\documentclass[12pt]{article}%

\RequirePackage[colorlinks,citecolor=blue,urlcolor=blue]{hyperref}
\usepackage[lmargin=25mm,rmargin=35mm,tmargin=35mm, bmargin=35mm]{geometry}
\usepackage{amsfonts}
\usepackage{amsmath}
\usepackage[english]{babel}
\usepackage{inputenc}
\usepackage[round]{natbib}
\usepackage{geometry}
\usepackage{float}
\usepackage{caption}
\usepackage{subcaption}
\usepackage{amsthm}
\usepackage{animate}
\usepackage{graphicx}
\usepackage{color}
\usepackage{ulem}

\graphicspath{{./Images/}}

\usepackage{algorithm}
\usepackage{algpseudocode}
\usepackage{pifont}

\restylefloat{table}
\def\S{\mathbb{S}}
\def\R{\mathbb{R}}

\def\bZ{\boldsymbol{Z}}
\def\bX{\boldsymbol{X}}

\def\bI{\boldsymbol{I}}
\def\bF{\boldsymbol{F}}

\def\bSigma{\pmb\Sigma}
\def\bLambda{\pmb\Lambda}

\def\bPsi{\pmb\Psi}
\def\bUpsilon{\pmb\Upsilon}

\begin{document}

\baselineskip=28pt \vskip 5mm
\begin{center} {\LARGE{\bf Fast and exact simulation of isotropic Gaussian random fields on $\mathbb{S}^{2}$ and $\S^{2}\times \R$}}
\end{center}

\vspace{0.5cm}

\begin{center}\large

Francisco Cuevas \footnote{\baselineskip=2pt Department of Mathematical Sciences, Aalborg University, Skjernvej 4A, 9220 Aalborg {\O}st  , Denmark }, Denis Allard \footnote{\baselineskip=2pt Biostatistics and Spatial Processes (BioSP), INRA PACA,  84914 Avignon, France}  and
Emilio Porcu \footnote{\baselineskip=2pt
School of Mathematics, Statistics and Physics, Newcastle University, NE1 7RU Newcaste Upon Tyne, UK.\\
Department of mathematics, University of Atacama. 
}
\end{center}

\begin{abstract}
We provide a method for fast and exact simulation of Gaussian random fields on spheres having isotropic covariance functions. The method proposed is then extended to Gaussian random fields defined over spheres cross time and having covariance functions that depend on geodesic distance in space and on temporal separation. 
The crux of the method is in the use of block circulant matrices obtained working on regular grids defined over longitude $\times$ latitude. 

\bigskip

\end{abstract}

\newpage
\section{Introduction}
Simulation of Gaussian random fields (GRFs) is important for the use of Monte Carlo techniques. Considerable work has been done to simulate GRFs defined over the $d$-dimensional Euclidean space, $\R^{d},$ with isotropic covariance functions. The reader is referred to \citet{Wood1994}, \citet{dietrich1997fast}, \citet{gneiting2006fast} and \citet{park2015block} with the references therein. See also \citet{emery2016improved} for extensions to the anisotropic and non stationary cases.
Yet, the literature on GRFs defined over two dimensional spheres (or just sphere) or spheres cross time has been sparse. Indeed, only few simulation methods for random fields on the sphere can be found in the literature. Amongst them, Cholesky descomposition and Karhunen-Lo\`eve expansion \citep{lang2015isotropic}. More recently, \citet{creasey2018fast} proposed an algorithm that decomposes a random field into one dimensional GRFs. Simulations of the 1d process are performed along with their derivatives, which are then transformed to an isotropic Gaussian random field on the sphere by Fast Fourier Transform (FFT). Following \citet{Wood1994}, Cholesky decomposition is considered as an \textit{exact method}, that is, the simulated GRF follows an exact multivariate Gaussian distribution. Simulation based on Karhunen-Lo\`eve expansion or Markov random fields are considered as \textit{approximated methods}, because the simulated GRF follows an approximation of the  multivariate Gaussian distribution \citep[see][]{lang2015isotropic,moller2015determinantal}. Extensions  to the spatially isotropic and temporally stationary GRF on the sphere cross time using space-time Karhunen-Lo\`eve expansion was considered in \citet{clarke2016regularity}.

It is well known that the computational cost to simulate a random vector at $n$ space-time locations using the Cholesky decomposition is $\mathcal{O}(n^{3}),$ which is prohibitively expensive for large values of $n$. Karhunen-Lo\`eve expansion requires the computation of Mercer coefficients of the covariance function \citep{lang2015isotropic,clarke2016regularity} which are rarely known. Finally, the method proposed by \citet{creasey2018fast} is restricted to a special case of spectral decomposition, which makes the method lacking generality. 

One way to reduce the computational burden is through the relationship between torus-wrapping, block circulant matrices and FFT. The use of this relationship has been introduced by \citet{Wood1994} and \citet{dietrich1997fast} when developing circulant embedding methods to simulate isotropic GRFs over regular grids of $\R^{d}$. Regular polyhedras are good candidates to be used as meshes for the sphere. However, on the sphere there exists only five regular polyhedras (the Platonic solids) limiting the number of regular points on the sphere to at most 30 \citep{coxeter1973regular}.

For a GRF on the sphere, using spherical coordinates, \citet{jun2008nonstationary} make use of circulant matrices to compute the exact likelihood of non-stationary covariance functions when the GRF is observed on a regular grid over $(0,\pi)\times(0,2\pi)$.

The aim of this paper is to develop a fast and exact simulation method for isotropic GRFs on the sphere and for spatially isotropic and temporally stationary GRFs on the sphere cross time. The proposed method requires an isotropic covariance function on the sphere, or a spatially isotropic and temporally stationary covariance function on the sphere cross time. One of the advantages of this method is the huge reduction of the computational cost to $\mathcal{O}(n\log(n))$. 


The paper is organized as follows. Section 2 details how to obtain circulant matrices on the sphere for isotropic covariance functions. In Section 3 we extend the simulation method to the case of the sphere cross time for spatially isotropic and temporally stationary covariance functions. The algorithms are detailed in Section 4 and in Section 5 we provide a simulation study. Finally, Section 6 contains some concluding remarks.

\section{Circulant Matrices over two dimensional spheres}
\label{sec:spheres}
Let $\S^{2} = \{s \in \R^{3}: \|s\| = 1\} \subset \R^{3}$ be the unit sphere centered at the origin,  equipped with the geodesic distance $\theta(s_1,s_2) := \arccos(\langle s_1,s_2 \rangle)$, for $s_1,s_2 \in \S^{2}$. We propose a new approach to simulate a finite dimensional realization from a real valued and zero mean, stationary, geodesically isotropic GRF $\textbf{X} = \{X(s): s \in \S^{2}\}$ with a given covariance function $$R(s_1,s_2) = \mathbb{E} \left[ Z(s_1) Z(s_2) \right] = r(\theta(s_1,s_2)), \quad s_1,s_2 \in \S^2. $$ 
Through the paper we equivalently refer to $R$ or $r$ as the covariance function of $\textbf{X}$. 
A list of isotropic covariance functions is provided in \cite{gneiting2013strictly}. In what follows, we use the shortcut $\theta$ for $\theta(s_1,s_2)$ whenever there is no confusion. 

\begin{figure}[h]
\centering
\includegraphics[scale=2]{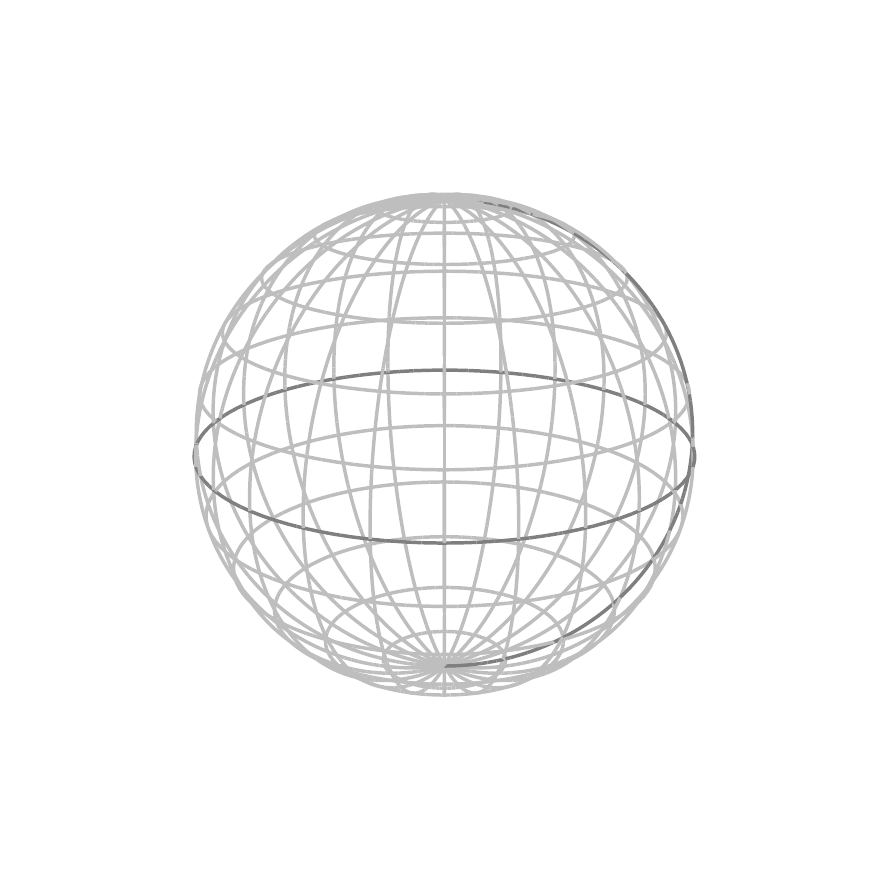} 
\caption{Example of the grid with N=24 and M=9 .\label{image1}}
\end{figure}

For a stationary isotropic random field on $\S^{2},$ the covariance functions and variogram are uniquely determined through the relation  \citep{huang2011validity}
\begin{equation} \label{variogram}
\gamma(\theta) = \sigma^{2} \Big( 1 - r(\theta) \Big ),  \quad 0\leq \theta \leq \pi.
\end{equation}
The basic requirements to simulate a GRF with the proposed method, are a grid, being regular over both longitude and latitude, and the computation of the covariance over this grid. 
For two integers $M,N \geq 2$ let $I = \{1,\ldots,N \}$ and $J = \{1,\ldots,M \},$  and define $\lambda_{i} = 2\pi i/N$ and $\phi_{j} = \pi j/M$ for $i \in I$ and $j \in J $ respectively. In the following, $s_{ij}=(\lambda_i,\phi_j)$ will denote the longitude--latitude coordinates of the point $s_{ij} \in \S^{2}$ and the set $\Omega_{MN} = \{(\lambda_{i},\phi_{j}): i \in I, j \in J \}$ defines a regular grid over $\S^{2}$ (see Figure \ref{image1}). The Cartesian coordinates of $s_{ij}$, expressed in $\R^{3}$, are
\begin{eqnarray}
s_{ij} = (x_{ij},y_{ij},z_{ij}) = (\cos \lambda_{i} \sin \phi_{j} , \sin \lambda_{i} \sin \phi_{j}, \cos \phi_{j}).
\end{eqnarray}
\noindent  Let us now define the random vector 
$$\textbf{X}_{i} = [X(s_{i1}), X(s_{i2}), \ldots, X(s_{iM})],$$ 
\noindent and the random field restricted to $\Omega_{MN}$ by $\textbf{X}_\Omega = [\textbf{X}_{1}, \textbf{X}_{2}, \ldots, \textbf{X}_{N}]$. The  matrix $\bSigma = \mbox{Var}[\textbf{X}_{\Omega}]$ has a block structure
\begin{eqnarray}
\bSigma = \left[
\begin{array}{cccc}
\bSigma_{1,1}  & \bSigma_{1,2} & \ldots & \bSigma_{1,N}\\
\bSigma_{1,2}  & \bSigma_{2,2} & \ldots & \bSigma_{2,N-1}\\
\vdots & \vdots & \ddots &  \vdots \\
\bSigma_{1,N}& \bSigma_{2,N} & \cdots & \bSigma_{N,N} 
\end{array}
\right],
\end{eqnarray}
where $\bSigma_{i,j} = \mbox{cov}(\textbf{X}_{i},\textbf{X}_{j}) = \bSigma_{j,i} $. 
Moreover, the geodesic distance can be written as
\begin{align}\label{polararclength}
\theta(s_{ik},s_{jl}) &= \arccos(x_{ik} x_{jl} + y_{ik} y_{jl} + z_{ik} z_{jl}) \notag\\
&= \arccos(\sin\phi_{k}\sin\phi_{l}(\cos\lambda_{i}\cos\lambda_{j} + \sin\lambda_{i}\sin\lambda_{j}) + \cos\phi_{k}\cos\phi_{l})\notag\\
&=\arccos(\sin\phi_{k}\sin\phi_{l}\cos(\lambda_{j} - \lambda_{i})) + \cos\phi_{k}\cos\phi_{l}).
\end{align}

\noindent Equation \eqref{polararclength} implies $\bSigma_{i,j} = \bSigma_{1,|j-i|+1}.$ Thus, writing $\bSigma_{i} = \bSigma_{1,i}$, we get 
\begin{eqnarray}
\bSigma 
= \left[
\begin{array}{cccc}
\bSigma_{1}  & \bSigma_{2} & \ldots & \bSigma_{N}\\
\bSigma_{N}  & \bSigma_{1} & \ldots & \bSigma_{N-1}\\
\vdots & \vdots & \ddots &  \vdots \\
\bSigma_{2}& \bSigma_{3} & \cdots & \bSigma_{1} 
\end{array}
\right].\label{BCMatrix}
\end{eqnarray}

Equation \eqref{BCMatrix} shows that $\bSigma$ is a \textit{symmetric block circulant matrix} \citep{davis2012circulant} 
related to the discrete Fourier transform as follows. Let $\bI_{M}$ be the identity matrix of order $M$ and $\bF_{N}$ be the Fourier matrix of order $N$, that is, $[\bF_{N}]_{i,k} = w^{(i-1)(k-1)}$ for $1\leq i,k \leq N$ where $w = e^{-2\pi \imath/N}$ and $\imath = \sqrt{-1}$. Following \citet{zhihao1990note}, the matrix $\bSigma$ is unitary block diagonalizable by $\bF_{N} \otimes \bI_{M}$, where $\otimes$ is the Kronecker product. Then, there exists $N$ matrices $\bLambda_{i}$, with $i\in I$, having dimension $M\times M,$ such that 
\begin{align}\label{fourierdecomposition}
\bSigma = \frac{1}{N} (\bF_{N} \otimes \bI_{M}) \bLambda (\bF_{N} \otimes \bI_{M})^{*}, \mbox{ with } 
\bLambda = 
\left[
\begin{array}{cccc}
\bLambda_{1}& \textbf{0} & \ldots& \textbf{0} \\
\textbf{0} &\bLambda_{2} & \ldots&\textbf{0} \\
\vdots& \vdots & \ddots&\vdots \\
\textbf{0} & \textbf{0} & \ldots&\bLambda_{N} \\
\end{array}
\right],
\end{align}

\noindent where $\boldsymbol{B}^{*}$ denotes the conjugate transpose of the matrix $\boldsymbol{B}$ and $\bf 0$ is a matrix of zeros of adequate size. The decomposition \eqref{fourierdecomposition} implies that the block matrix $\bLambda$ can be computed through the discrete Fourier transform of its first block row, that is,
\begin{align}
\left[
\begin{array}{cccc}
\bSigma_{1}& \bSigma_{2} &\cdots &\bSigma_{N}
\end{array}
\right](\bF_{N}\otimes \bI_{M}) 
= 
\left[
\begin{array}{cccc}
\bLambda_{1}& \bLambda_{2}& \cdots &\bLambda_{N}
\end{array}
\right]. \label{eq:Lambda}
\end{align}
Componentwise,  (\ref{eq:Lambda}) becomes
\begin{align}\label{LambdaElement}
\left[
\begin{array}{cccc}
\bSigma_{1}^{jl}&\bSigma_{2}^{jl}&\cdots&\bSigma_{N}^{jl}
\end{array}
\right]
\bF_{N} = 
\left[
\begin{array}{cccc}
\bLambda_{1}^{jl}&\bLambda_{2}^{jl}&\cdots&\bLambda_{N}^{jl}
\end{array}
\right],
\end{align}
where $j,l \in J.$  Since $\bSigma$ is positive definite (semi-definite), it is straightforward from the decomposition (\ref{fourierdecomposition}) that the matrix $\bLambda$ is positive definite (semi-definite), and thus that each matrix $\bLambda_i$  is also positive definite (semi-definite), for $i \in I$. Hence we get
\begin{align} \label{sqrtC}
\bSigma^{1/2} = \frac{1}{\sqrt{N}} (\bF_{N} \otimes \bI_{M}) \bLambda^{1/2},
\end{align}
which is what is needed for simulation. A simulation algorithm based on Equation (\ref{sqrtC}) is provided in Section \ref{algorithm}.




\section{Circulant embedding on the sphere cross time}
\label{sec:spheresxtime}
We now generalize this approach on the sphere cross time for a spatially isotropic and temporally stationary GRF $\textbf{X} = \{X(s,t) : (s,t) \in \S^{2}\times \R \}$ with zero mean and a given covariance function, $R$, defined as  

$$ R \Big ( (s_1,t_1),(s_2,t_2) \Big ) := r(\theta(s_1,s_2),|t_1-t_2|), \qquad (s_i,t_i) \in \S^2 \times \R, \quad i=1,2.$$ 

We analogously define the space-time stationary variogram $\gamma: [0,\pi] \times \mathbb{R} \mapsto \R$ as
\begin{equation}\label{st_variog}
\gamma(\theta, u) = \sigma^{2}(1 - r(\theta,u)), \qquad \theta \in 
[0,\pi], u \in \mathbb{R}.
\end{equation}
Let us denote $H$ the time horizon at which we wish to simulate. Let $T$ be a positive integer and define the regular time grid $t_{\tau} = \tau H/T$ with $\tau = \{1,\ldots,T\}$. Define the set $\Omega_{NMT} = \{(\lambda_{i},\phi_{j},t_{\tau}): i \in I, j\in J,\tau \in \{1,\dots,T\} \},$ and the random vectors
\textcolor{blue}{}
\begin{align}
\textbf{X}_{i,\tau} &= [X(s_{i1},t_{\tau}),~X(s_{i2},t_{\tau}),~\cdots,~X(s_{iM},t_\tau)], \notag\\
\textbf{X}_{\Omega,\tau} &= [\textbf{X}_{1,\tau},~\textbf{X}_{2,\tau},~\cdots,~\textbf{X}_{N,\tau}], \notag\\
\textbf{X}_{\Omega} &= [\textbf{X}_{\Omega,0},~\textbf{X}_{\Omega,1},~\cdots,~\textbf{X}_{\Omega,L}]. \notag
\end{align}

\noindent The associated covariance matrices are
\begin{align*}
\bPsi_{i,k}(\tau,\tau') &= \mbox{cov}(\textbf{X}_{i,\tau},\textbf{X}_{k,\tau'}),\notag\\
\bPsi(\tau,\tau') &= \mbox{cov}(\textbf{X}_{\Omega,\tau},\textbf{X}_{\Omega,\tau'}),\notag\\
\bPsi &= \mbox{Var}(\textbf{X}_\Omega),\notag
\end{align*}
with $i,k \in I$ and $1 \leq \tau,\tau' \leq T$.
\noindent The assumption of temporal stationarity implies that $\bPsi_{i,k}(\tau,\tau') = \bPsi_{i,k}(|\tau'-\tau|).$ Therefore, $\bPsi(\tau,\tau') = \bPsi(\tau',\tau) = \bPsi(|\tau'- \tau|).$ As a consequence, for fixed $\tau,\tau'$ the matrix $\bPsi(|\tau'-\tau|)$ is a block circulant matrix with dimension $NM\times NM$, however, $\bPsi$ is not. To tackle this problem we consider the torus-wrapped extension of the grid $\Omega_{MNL}$ over the time variable which is detailed as follows. First, let $\kappa$ be a positive integer and let $g: \R \mapsto \R$ be defined as
\begin{align}
g(\tau) = \left\{ \begin{array}{ccl}
\frac{\tau H}{T}& \mbox{ if } & 1 \leq \tau \leq \kappa T,\\
\\
\frac{(2 \kappa T - \tau)H}{T}& \mbox{ if } & \kappa T< \tau \leq (2\kappa T - 1).\\
\end{array}\right.
\end{align}
\noindent Then, the matrix

\begin{align} \label{Embedded_cov}
\tilde{\bPsi} & = \left[ \begin{array}{cccccc}
\bPsi(g(0))  & \bPsi(g(1)) & \ldots & \bPsi(g(\kappa T)) & \ldots & \bPsi(g(2\kappa T - 1))\\
\bPsi(g(1))  & \bPsi(g(0)) & \ddots & \bPsi(g(\kappa T-1)) & \ldots & \bPsi(g(2\kappa T - 2))\\
\vdots & \vdots & \vdots &  \vdots & \vdots & \vdots \\
\bPsi(g(1))& \bPsi(g(2)) & \cdots & \bPsi(g(\kappa T+1)) &  \ldots & \bPsi(g(0)) 
\end{array} \right] \notag\\
& & & & &  \notag \\
  &= \left[
 \begin{array}{cccccc}
 \bPsi(0)  & \bPsi(1) & \ldots & \bPsi(\kappa T) & \ldots & \bPsi(1)\\
 \bPsi(1)  & \bPsi(0) & \ddots & \bPsi(\kappa T-1) &  \ldots & \bPsi(2)\\
 \vdots & \vdots & \vdots &  \vdots & \vdots & \ddots \\
 \bPsi(1)& \bPsi(2) & \cdots & \bPsi(\kappa T-1) & \ldots & \bPsi(0) 
 \end{array}
 \right]
\end{align}

\noindent is a $(NM(2\kappa T - 1))\times(NM(2\kappa T - 1))$ block circulant matrix. Considering that each $\bPsi(g(\tau))$ is a block circulant matrix for $0 \leq (\tau \leq 2\kappa T - 1)$ and using decomposition \eqref{fourierdecomposition} we get 
\begin{align}\label{DoubleFourier}
\tilde{\bPsi} = [\bI_{2\kappa L - 1}\otimes (\bF_{N} \otimes \bI_{M})][\bF_{2\kappa L - 1}\otimes \bI_{MN}] \bUpsilon [\bI_{2\kappa L - 1}\otimes (\bF_{N} \otimes \bI_{M})]^{*}[\bF_{2\kappa L - 1}\otimes \bI_{MN}]^{*},
\end{align}
\noindent with
$$\bUpsilon = \left[\begin{array}{cccc}
\bUpsilon_{0}&\textbf{0} &\cdots &\textbf{0}\\
\textbf{0}&\bUpsilon_{1} &\cdots &\textbf{0}\\
\vdots&\vdots & \ddots &\vdots\\
\textbf{0}&\textbf{0} &\cdots & \bUpsilon_{(2\kappa T - 1)N }
\end{array}\right],$$
where $\bUpsilon_{i}$ is a $M\times M$ matrix for all $i \in I$. This decomposition allows to compute the square root of $\tilde{\bPsi}$ using the FFT algorithm twice. However, this procedure does not ensure $\tilde{\bPsi}$ to be a positive definite matrix. To circumvent this problem, \citet{Wood1994} propose to choose a large value of $\kappa$, at the cost of an increasing  computational burden \citep{gneiting2006fast}. 

\section{Simulation algorithms}\label{algorithm}
Sections~\ref{sec:spheres} and~\ref{sec:spheresxtime} provided the mathematical background for simulating GRFs on a regular (longitude, latitude) grid based on circulant embedding matrices. Algorithms~\ref{algo:sphere} and~\ref{algo:spherextime} presented in this section detail the procedures to to be implemented for simulating on $\S^{2}$ and $\S^{2}\times \R$ respectively. The suggested procedure is fast to compute and requires moderate memory storage since only blocks $\bSigma_{i}$ (resp.  $\bPsi(i)$) and their square roots, of size $M \times M$ each, need to be stored and computed. FFT algorithm is used to compute the matrices $\bLambda$ and $\bUpsilon$ through Equation \eqref{DoubleFourier}. The last step of both Algorithms can be calculated using FFT, generating a complex-valued vector $\textbf{Y}$ where the real and imaginary parts are independent. In addition, both algorithms can be parallelized. Also, a small value of $\kappa$ means that less memory is required to compute Algorithm 2.

Some comments about the computation of $\bLambda^{1/2}$ and $\bUpsilon^{1/2}$ are worth to mention. For a random field over $\S^{2}$ or $\S^{2} \times \R,$ the matrices $\bLambda$ and $\bUpsilon$ can be obtained through Cholesky decomposition if the underlying covariance function $r$ is strictly positive definite on the appropriate space. In the case of a positive semi-definite covariance function, both matrices become ill-conditioned, thus, generalized inverse must be used.  

\begin{algorithm}[ht]
\caption{\label{algo:sphere} Circulant embedding algorithm to simulate a GRF on $\S^{2}$}
\begin{algorithmic}
\Require Integers $M, N$; space locations $\{ s_{ij}), i\in I, j \in J\}$; stationary, isotropic, covariance function $r(\cdot)$
\\
\For{$i \in I$}
\State Build the matrices $\bSigma_{i} = \hbox{cov}(\bX_1,\bX_i)$ with elements 
$[\bSigma_{i}]_{j,\ell} = r(\arccos(\langle s_{1j},s_{i,\ell} \rangle))$,  $1 \leq j, \ell \leq M$
\State Compute the matrices $\bLambda_{i}$ using \eqref{LambdaElement}
\EndFor
\State Compute $\bLambda^{1/2}= \hbox{diag}(\bLambda^{1/2}_1,\dots,\bLambda^{1/2}_N)$
\State Generate a sample from $\textbf{Z} \sim N(0,\bI_{MN})$
\State Compute  $\textbf{Y}=\frac{1}{\sqrt{N}}(\bF_{N} \otimes \bI_{M}) \bLambda^{1/2}\bZ$
\State {\bf return}  $\textbf{Y}$
\end{algorithmic}
\end{algorithm}

\begin{algorithm}[ht]
\caption{\label{algo:spherextime} Circulant embedding algorithm to simulate a GRF on $\S^{2}\times \S^{2} \times \R$}
\begin{algorithmic}
\Require Integers $M, N, T, \kappa$, space-time locations $\{ s_{ijt_\tau}), i\in I, j \in J, \tau \in \{1,\dots,T\} \}$;  stationary and isotropic in space and stationary in time covariance function $r(\cdot,\cdot)$ 
\\
\For{$\tau \in \{1,\ldots,(2\kappa T-1)N\}$} 
\State Compute the matrices $\bPsi(g(\tau))$ with block matrices $[\bPsi_{i,j}(g(\tau))]$ with elements $[\bPsi_{i,j}(g(\tau))]_{k,l} = r(\arccos(\langle s_{ik}, s_{jl} \rangle),g(\tau))$
\State Compute the matrices $\bUpsilon_{\tau}$ using \eqref{DoubleFourier}
\EndFor
\State Compute $\bUpsilon^{1/2} = \mbox{diag}(\bUpsilon^{1/2}_{0},\ldots,\bUpsilon^{1/2}_{(2\kappa T - 1)N})$
\State Generate a sample from $\textbf{Z} \sim N(0,\bI_{MN(2\kappa L - 1)})$
\State Compute $\textbf{Y}= \frac{1}{\sqrt{N}}[\bI_{2\kappa L - 1}\otimes (\bF_{N} \otimes \bI_{M})][\bF_{2\kappa L - 1}\otimes \bI_{MN}]\bUpsilon^{1/2}\bZ$
\State {\bf return}  $\textbf{Y}$
\end{algorithmic}
\end{algorithm}

\clearpage
 
\section{Simulations}\label{results}

\begin{figure}[h]
\centering
\includegraphics[scale=0.7]{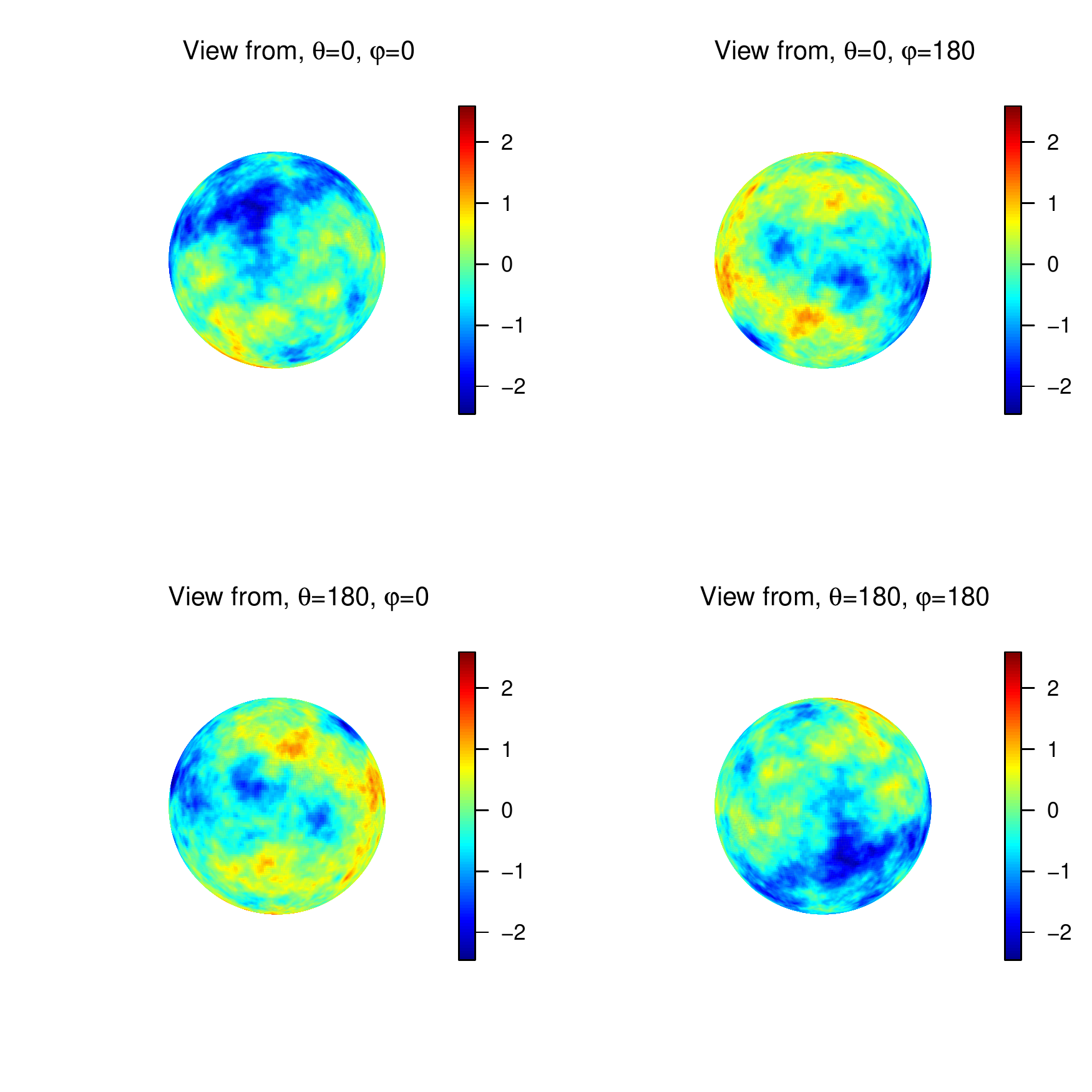} 
\caption{Realization of a GRF with covariance function given by Equation \eqref{exponential_cova}. The parameters of the simulations are $M=180$,$N=360$.\label{image2}}
\end{figure}

Through this section we assume that $r(0) = 1$ and $r(0,0) = 1$ for covariance functions in $\S^{2}$ and $\S^{2}\times\R$ respectively. To illustrate the speed and accuracy of Algorithm~\ref{algo:sphere}, we compare it with Cholesky and eigenvalue decompositions, to obtain a square root of the covariance matrix. Because Algorithm~\ref{algo:sphere} only needs the first block row of the covariance matrix, a fair comparison is made by measuring the calculation after the covariance matrix was calculated. The covariance model used in this simulation is the exponential covariance function \citep{chiles1999geoestadistics}, defined as  

\begin{equation}\label{exponential_cova}
r_{0}(\theta) = \exp\left(-\frac{\theta}{\phi_{0}}\right), \qquad \theta \in [0,\pi],
\end{equation}

\noindent where $\phi_{0} > 0$ has been chosen to ensure that $r(\pi/2) = 0.05.$ Table \ref{TableTime} shows the computational time (in seconds) needed for each method to generate the GRF on each grid. Reported times are based on a server with 32 cores 2x Intel Xeon e5‑2630v3, 2.4 Ghz processor and 32 GB RAM. Algorithm~\ref{algo:sphere} is always faster. For the largest mesh, the eigenvalue and the Cholesky decomposition methods do not work because of storage problems. An example of a realization of the GRF is showed in Figure \ref{image2}. 
\begin{table}[h]
\centering
\begin{tabular}{c|rll}
Parameters&Circ. Embed. & Cholesky & Eigen\\ \hline
N=18, M=6&0.016 & 0.008 & 0.032\\ \hline
N=40, M=13&0.021&0.541&1.047\\ \hline
N=60, M=20&0.057&2.397&5.832 \\ \hline
N=120, M=40&0.121&24.781&278.228 \\ \hline
N=360, M=180 &16.433&--&-- \\ \hline
\end{tabular}
\caption{Time (in seconds) needed for each algorithm to be completed. When $N=360$ and $M=180$ the Cholesky and the eigenvalue decompositions do not work because the computer is not able to storage the covariance matrix. Results were based in a server with 32 cores 2x Intel Xeon e5‑2630v3, 2.4 Ghz processor and 32 GB RAM}
\label{TableTime} 
\end{table}

To study the simulation accuracy, we make use of variograms as defined through (\ref{variogram}). Specifically, we estimate the variogram nonparametrically through
\begin{equation}\label{variog_estimate}
\hat{\gamma}(\theta) = \frac{1}{|N_{l}(\theta)|} \sum_{i,j,i^{\prime},j^{\prime}}(X(s_{ij}) - X(s_{i^{\prime}j^{\prime}}))^{2}\mathbb{I}_{N_{l}(\theta)}(s_{ij},s_{i^{\prime}j^{\prime}}),
\end{equation}
\noindent where $\mathbb{I}_{A}(x)$ is the indicator function of the set $A$, $l$ is a bandwidth parameter and 
$$N_{l}(\theta) = \{(s_{1},s_{2}) \in \S^{2}: |\theta(s_{1},s_{2}) - \theta|\leq l\}.$$

\noindent We perform our simulations, using $M = 30, N = 60$ (that is, $n = 1800$), corresponding to a $6~\times~6$ degree regular longitude-latitude grid on the sphere, under three covariance models \citep{gneiting2013strictly}:

\begin{enumerate}
\item The exponential covariance function defined by \eqref{exponential_cova}
\item The generalized Cauchy model defined by
$$r_{1}(\theta) = \left(1+ \left(\frac{\theta}{\phi_{1}}\right)^{\alpha}\right)^{-\frac{\beta}{\alpha}}, \qquad \alpha \in (0,1], \quad \phi_{1}, \beta > 0$$

\item The Matern model defined as
$$r_{2}(\theta) = \frac{2^{1-\nu}}{\Gamma(\nu)} \left(\frac{\theta}{\phi_{2}}\right)^{\nu} K_{\nu}\left(\frac{\theta}{\phi_2}\right), \qquad \nu \in (0,1/2], \phi_{2} > 0$$
\end{enumerate}
\noindent where $K_{\nu}(\cdot)$ is the Bessel function of second kind of order $\nu$ and $\Gamma(\cdot)$ is the Gamma function. In this simulation study we use $\phi_0 = 0.5243$, $\alpha = 0.75$, $\beta = 2.5626$, $\phi_{1} = 1$, $\nu = 0.25$ and $\phi_{2} = 0.7079$. Such setting ensures that $r_{i}(\pi/2) = 0.05$ for $i=0,1,2$. 
Note that the regularity parameter is restricted to the interval $(0,1/2]$ to ensure positive definiteness on $\S^2$ \citep{gneiting2013strictly}.

\begin{figure}[h]
\begin{subfigure}{.32\linewidth}
\centering
\captionsetup{justification=centering}
\includegraphics[scale=0.32]{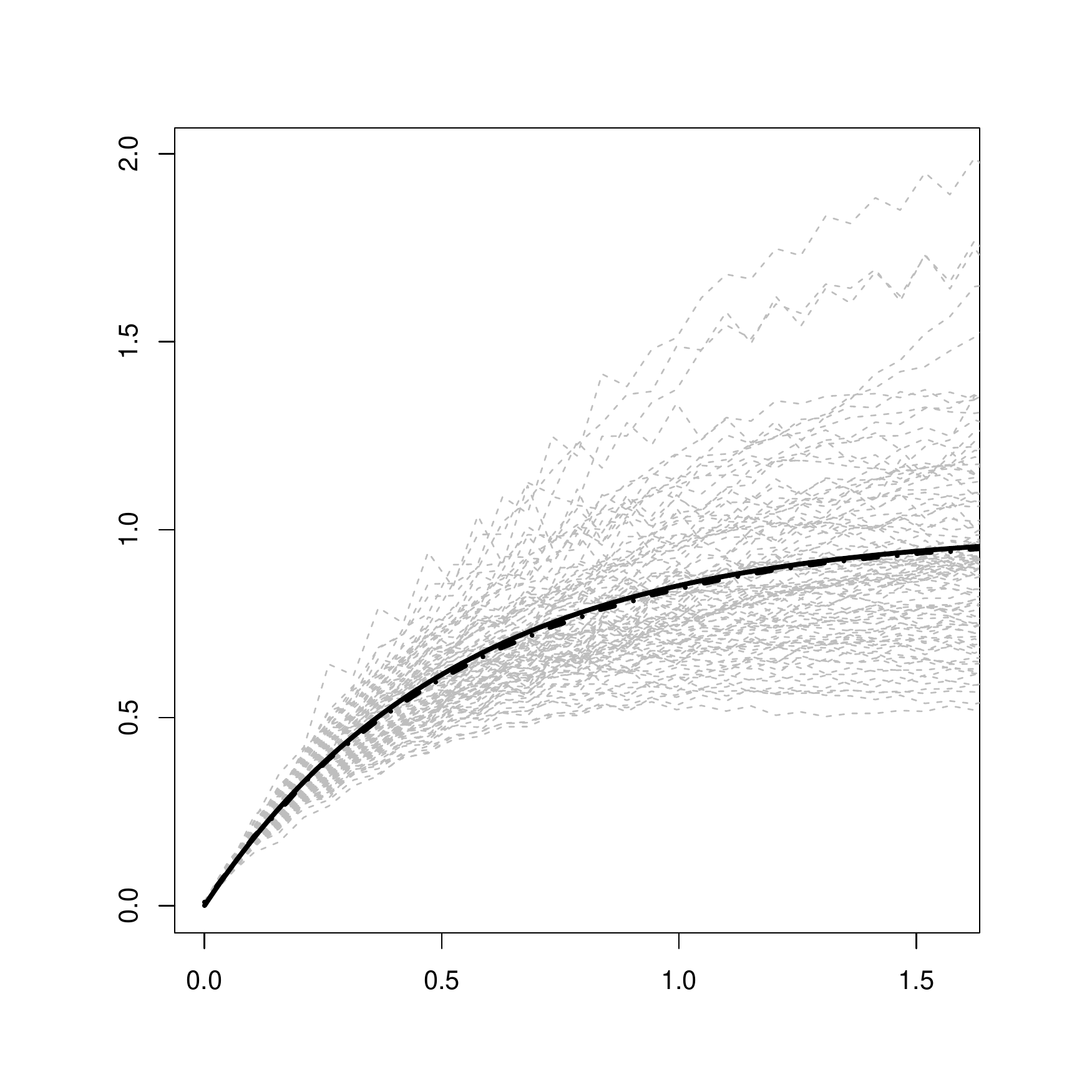} 
\caption{Exponential model}
\label{variogram_a}
\end{subfigure}
\begin{subfigure}{.32\linewidth}
\centering
\captionsetup{justification=centering}
\includegraphics[scale=0.32]{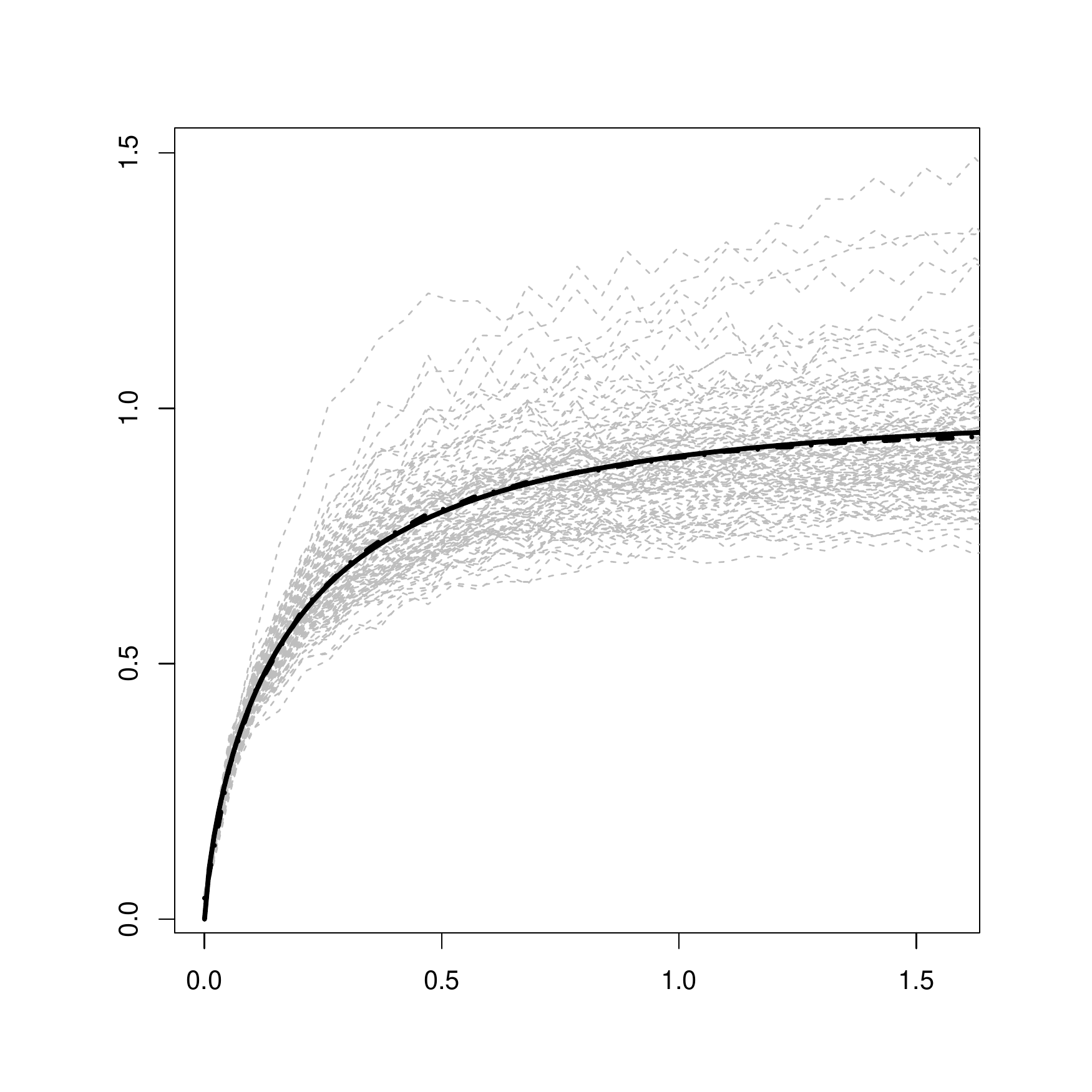} 
\caption{Generalized Cauchy}
\label{variogram_b}
\end{subfigure}
\begin{subfigure}{.32\linewidth}
\centering
\captionsetup{justification=centering}
\includegraphics[scale=0.32]{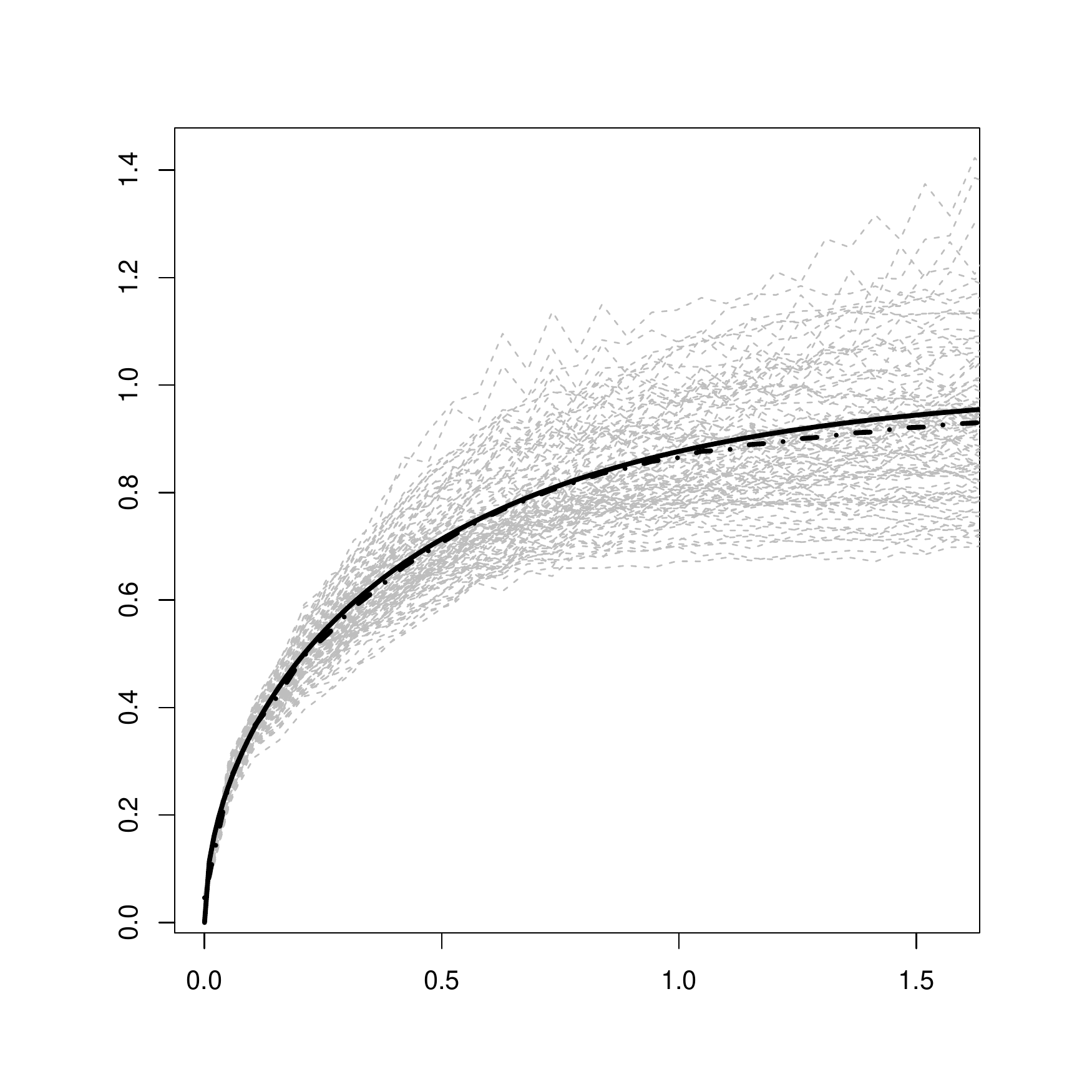} 
\caption{Matern model}\label{variogram_c}
\end{subfigure}
\\[1ex]
\begin{subfigure}{.32\linewidth}
\centering
\includegraphics[scale=0.3]{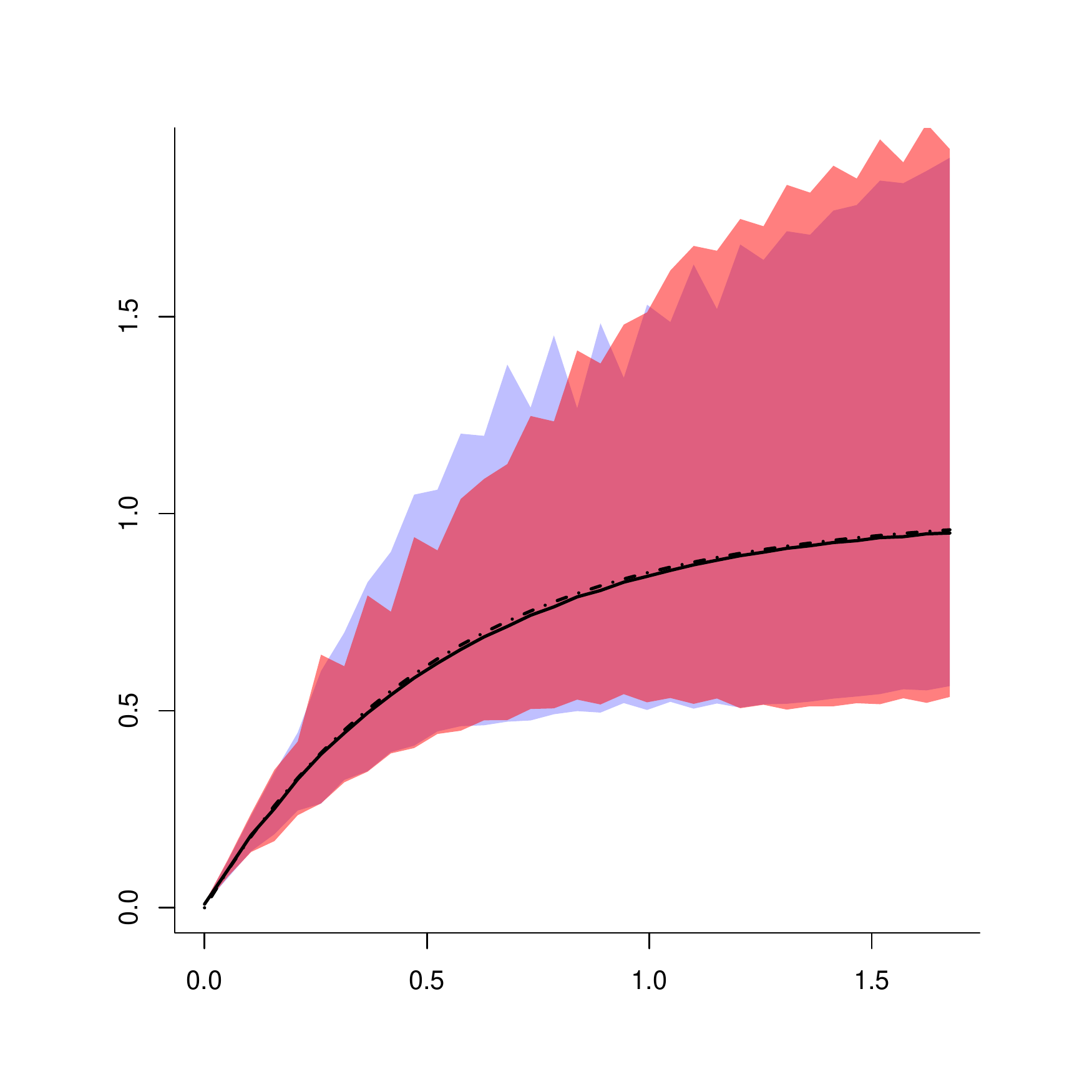} 
\caption{Envelope exponential}
\label{env_exponential}
\end{subfigure}
\hfill
\begin{subfigure}{.32\linewidth}
\centering
\includegraphics[scale=0.3]{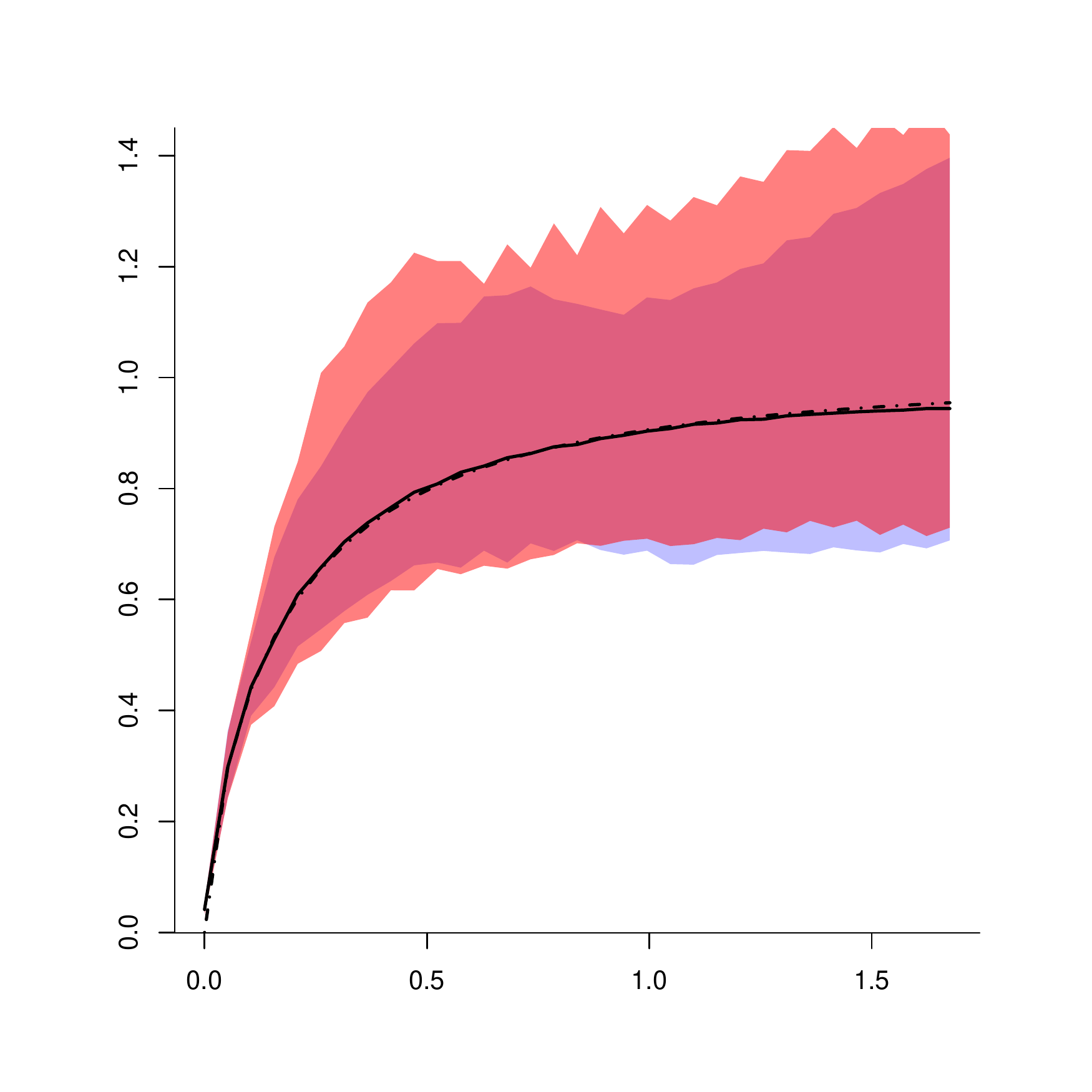} 
\caption{Envelope Cauchy}
\label{env_cauchy}
\end{subfigure}
\hfill
\begin{subfigure}{.32\linewidth}
\centering
\includegraphics[scale=0.3]{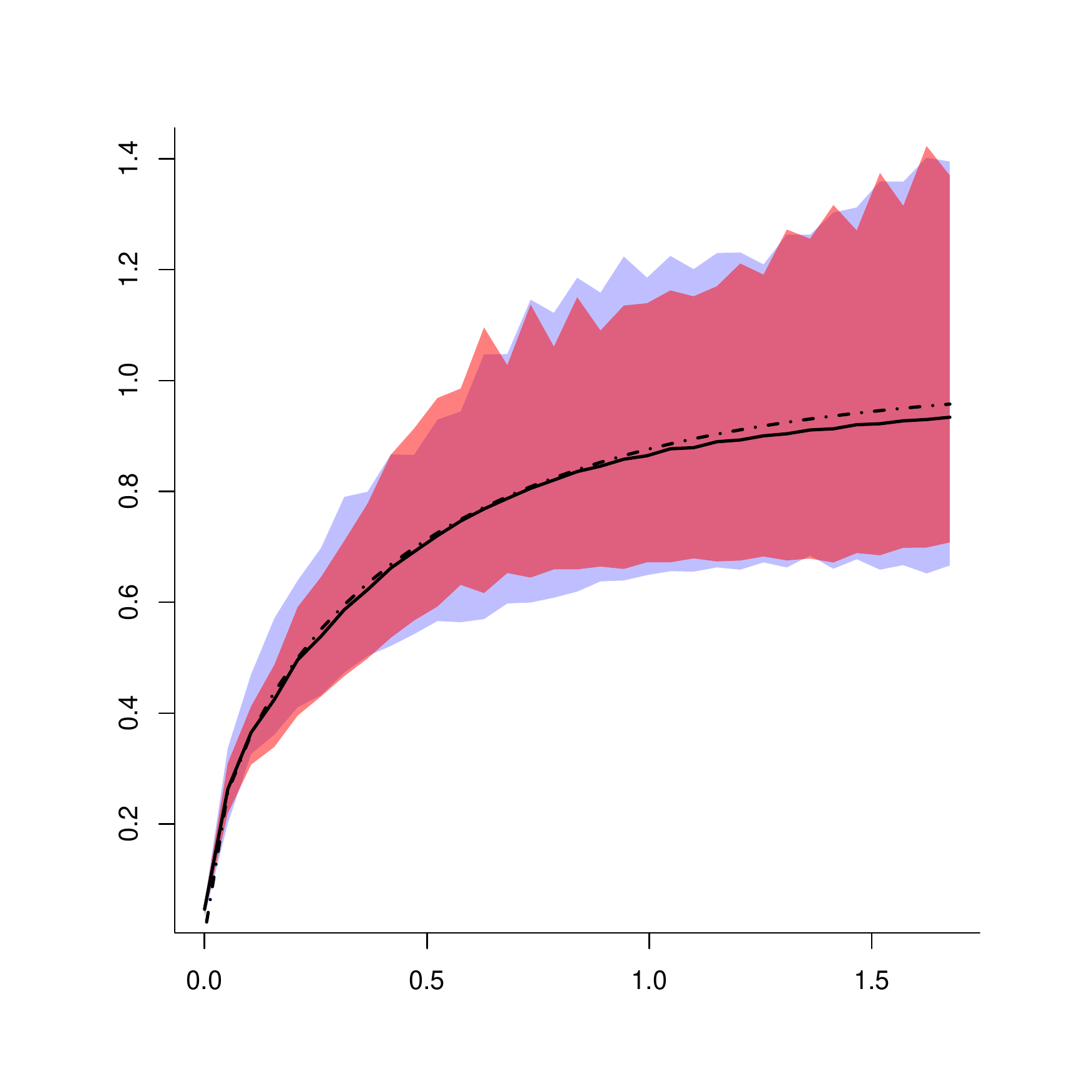} 
\caption{Envelope Matern}
\label{env_matern}
\end{subfigure}
\caption{Variogram estimates for $100$ simulations of different random fields with different models using circulant embedding. Panels \ref{variogram_a}--\ref{variogram_c} shows each realization and Panels (d)-(f) shows envelopes for simulations using Cholesky (blue envelope) and circulant embedding (red envelope). 
The black line is the true variogram function, gray lines are the estimated functions and black dotdashed points are the mean of the simulations.}\label{Spatial_variogram_1}
\end{figure}

For each simulation, Cholesky decomposition is used to compute $\Lambda^{1/2}$ and the variogram estimates in Equation \eqref{variog_estimate} are computed. 100 simulations have been performed for each covariance model using Algorithm \ref{algo:sphere} and using direct Cholesky decomposition for comparison. Figure \ref{variogram_a} -- \ref{variogram_c} show the estimated variogram for each simulation and the average variogram as well. Also, global rank envelopes \citep{myllymaki2017global} where computed for the variogram under each simulation algorithms. The average variograms match almost perfectly. The superimposition of the envelopes shows that our approach generates the same variability as the Cholesky decomposition. 


A similar simulation study is provided for a temporal stationary and spatially isotropic random field on $\S^{2}\times \R.$ We provide a non-parametric estimate of \eqref{st_variog} through

$$ \widehat{\gamma}(\theta,u) = \frac{1}{|N_{l,l^{\prime}}(\theta,u)|} \sum_{i,j,k,i^{\prime},j^{\prime},k^{\prime}} \Big (X(s_{ij},t_{k}) - X(s_{i^{\prime}j^{\prime}},t_{k^{\prime}}) \Big )^{2}\mathbb{I}_{N_{l,l^{\prime}}(\theta,u)}\Big ( (s_{ij},t_k), (s_{i^{\prime}j^{\prime}},t_{k^{\prime}}) \Big ),$$



\noindent where $\theta \in [0,\pi], u \in \R$ and 

$$N_{l,l^{\prime}}(\theta,u) = \{(s_{i},t_i) \in \S^{2}\times\ \R, i=1,2  \; : |\theta(s_{1},s_{2}) - \theta |\leq l, |t_1 - t_2| - u|\leq l^{\prime}\}.$$



\begin{figure}[h]
\begin{subfigure}{.32\linewidth}
\centering
\includegraphics[scale=0.3]{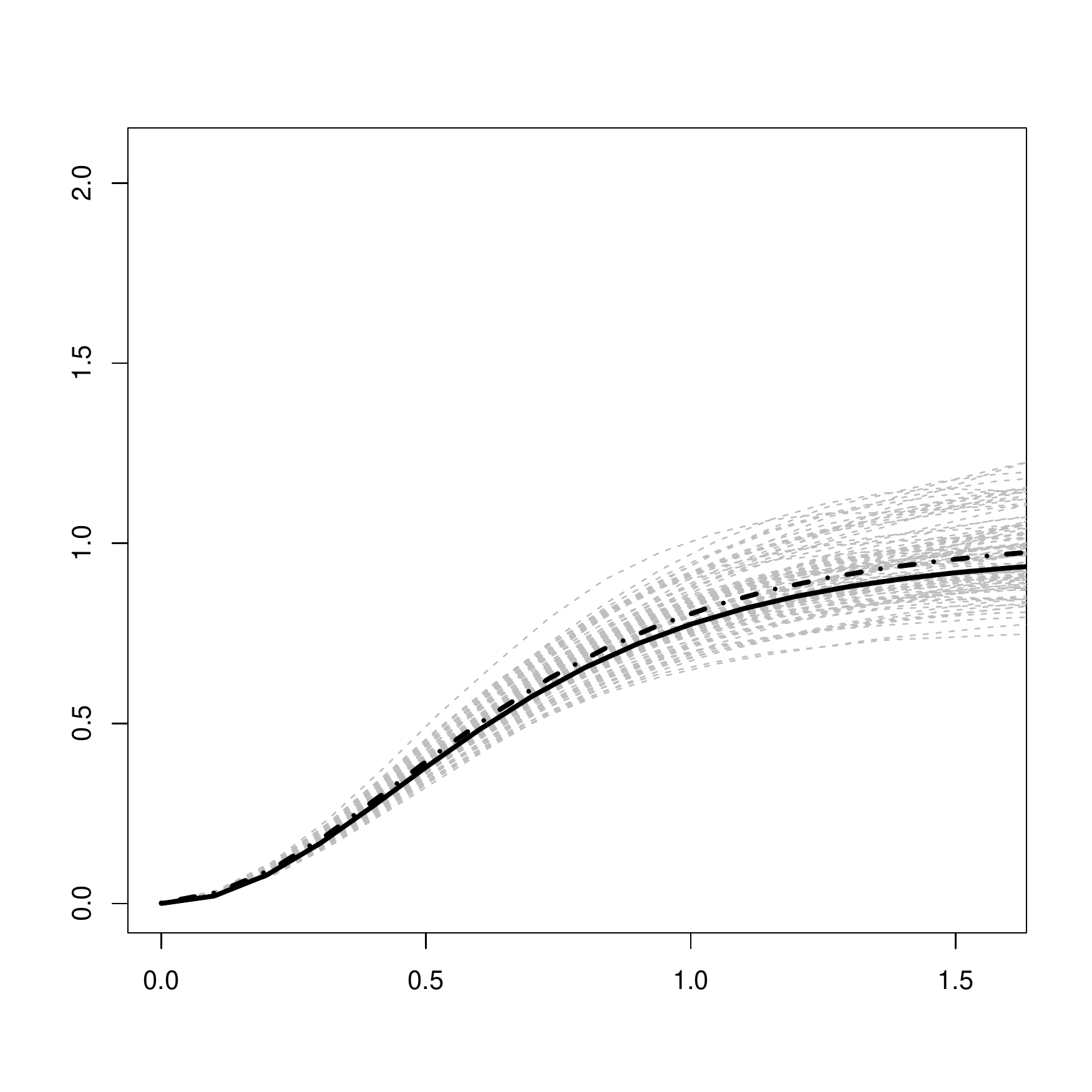} 
\caption{Spatial margin}
\label{exp_sm}
\end{subfigure}
\hfill
\begin{subfigure}{.32\linewidth}
\centering
\includegraphics[scale=0.3]{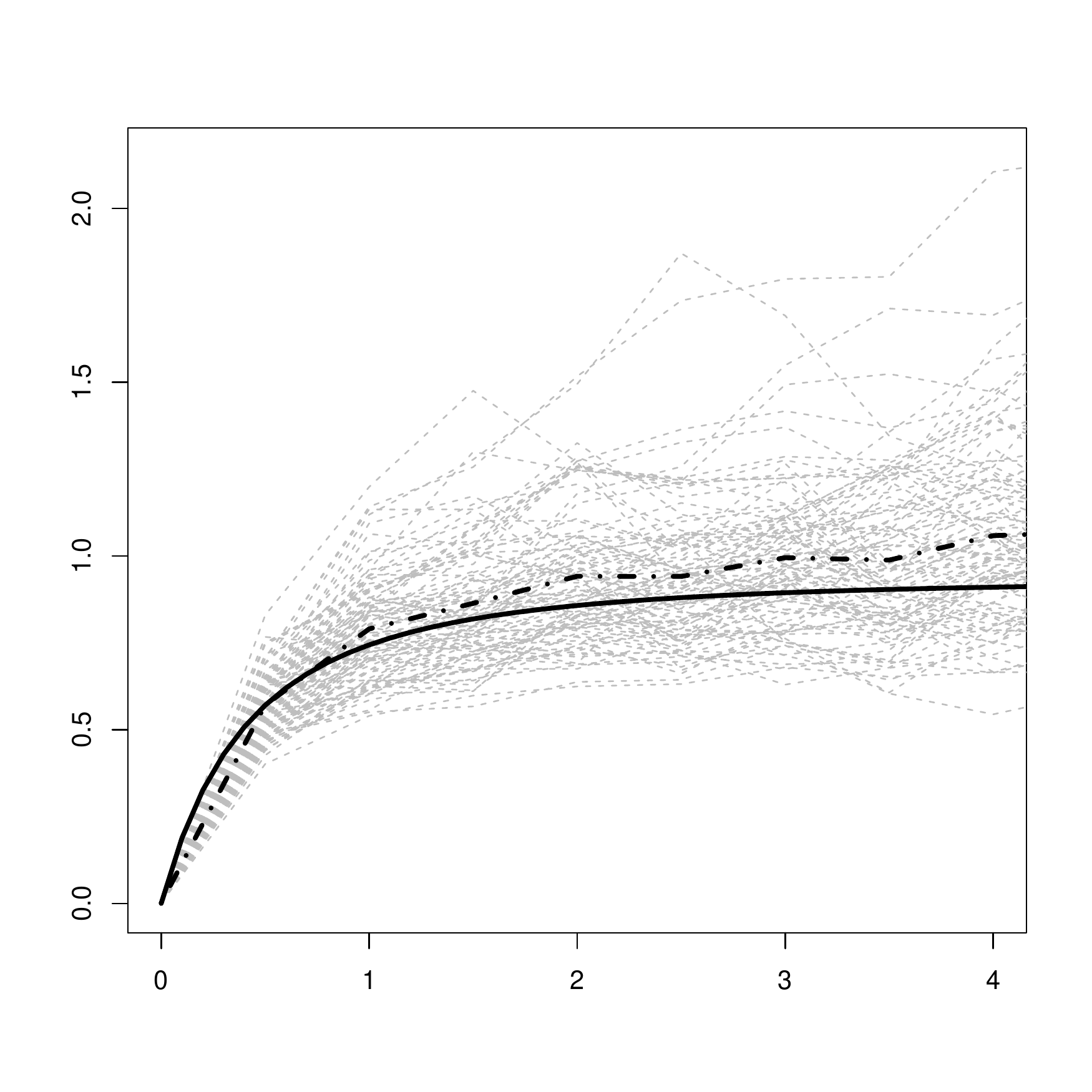} 
\caption{Temporal margin}
\label{exp_tm}
\end{subfigure}
\hfill
\begin{subfigure}{.32\linewidth}
\centering
\includegraphics[scale=0.3]{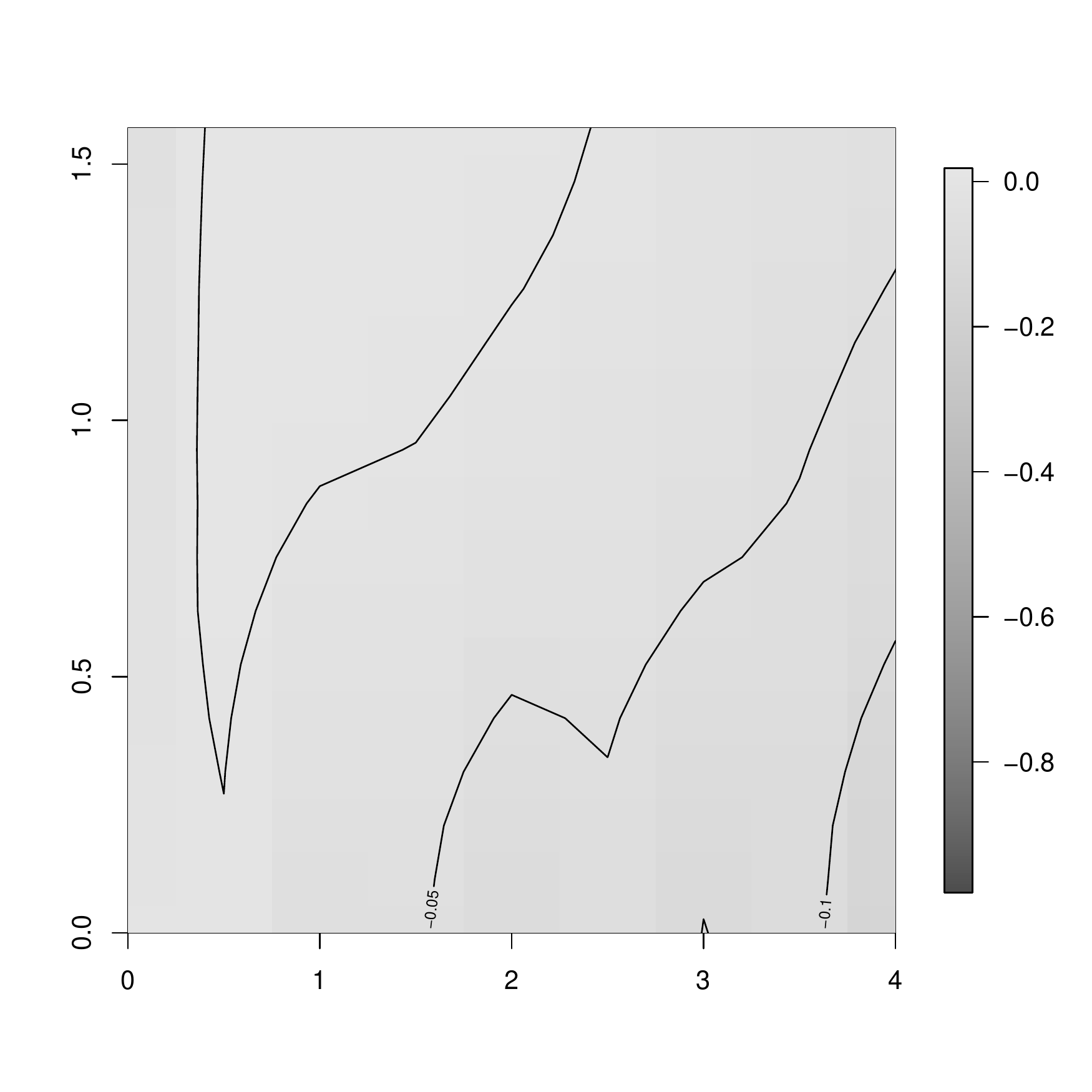} 
\caption{Variogram residuals}
\label{exp_vr}
\end{subfigure}
\\[1ex]
\begin{subfigure}{.32\linewidth}
\centering
\includegraphics[scale=0.3]{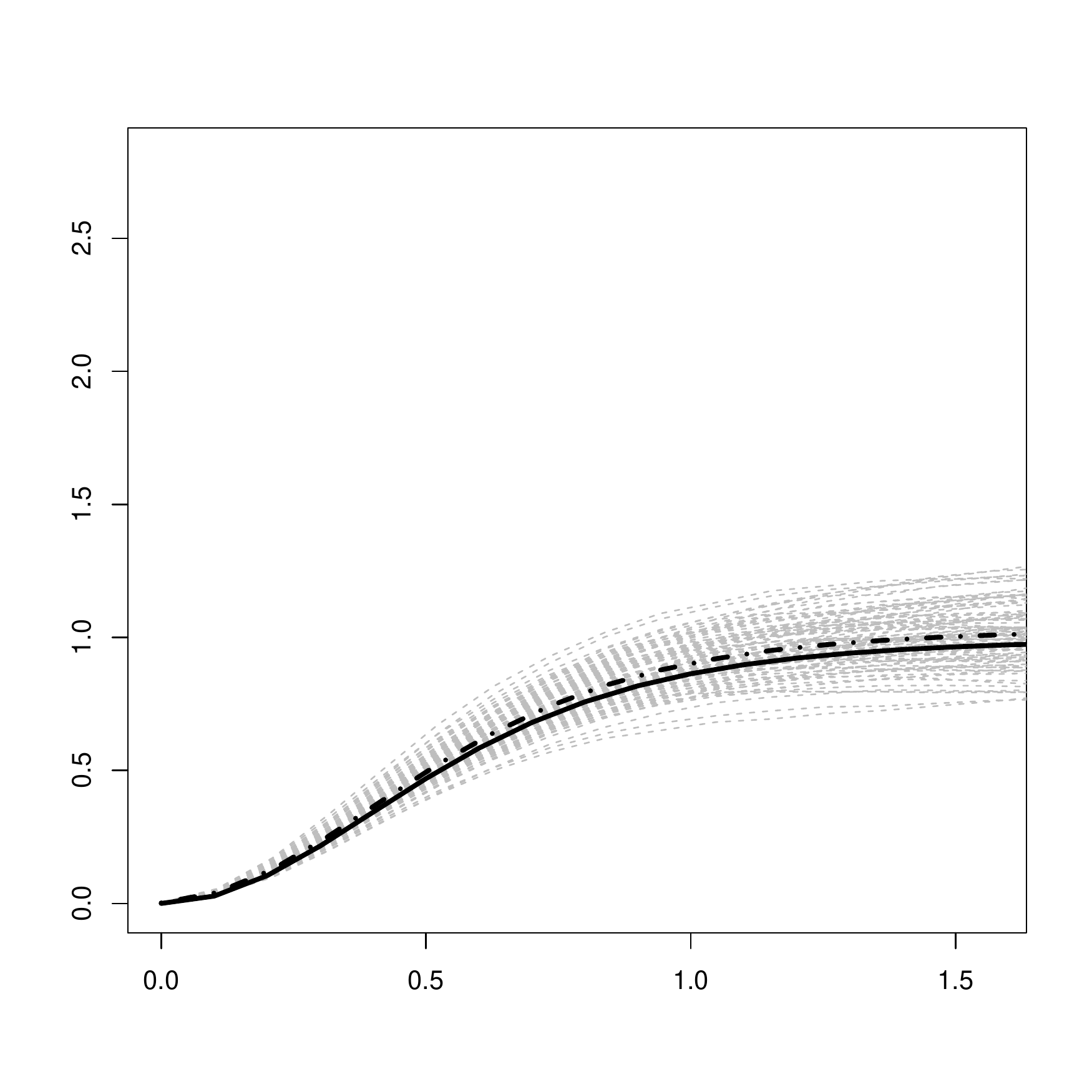} 
\caption{Spatial margin}
\label{gen_sm}
\end{subfigure}
\hfill
\begin{subfigure}{.32\linewidth}
\centering
\includegraphics[scale=0.3]{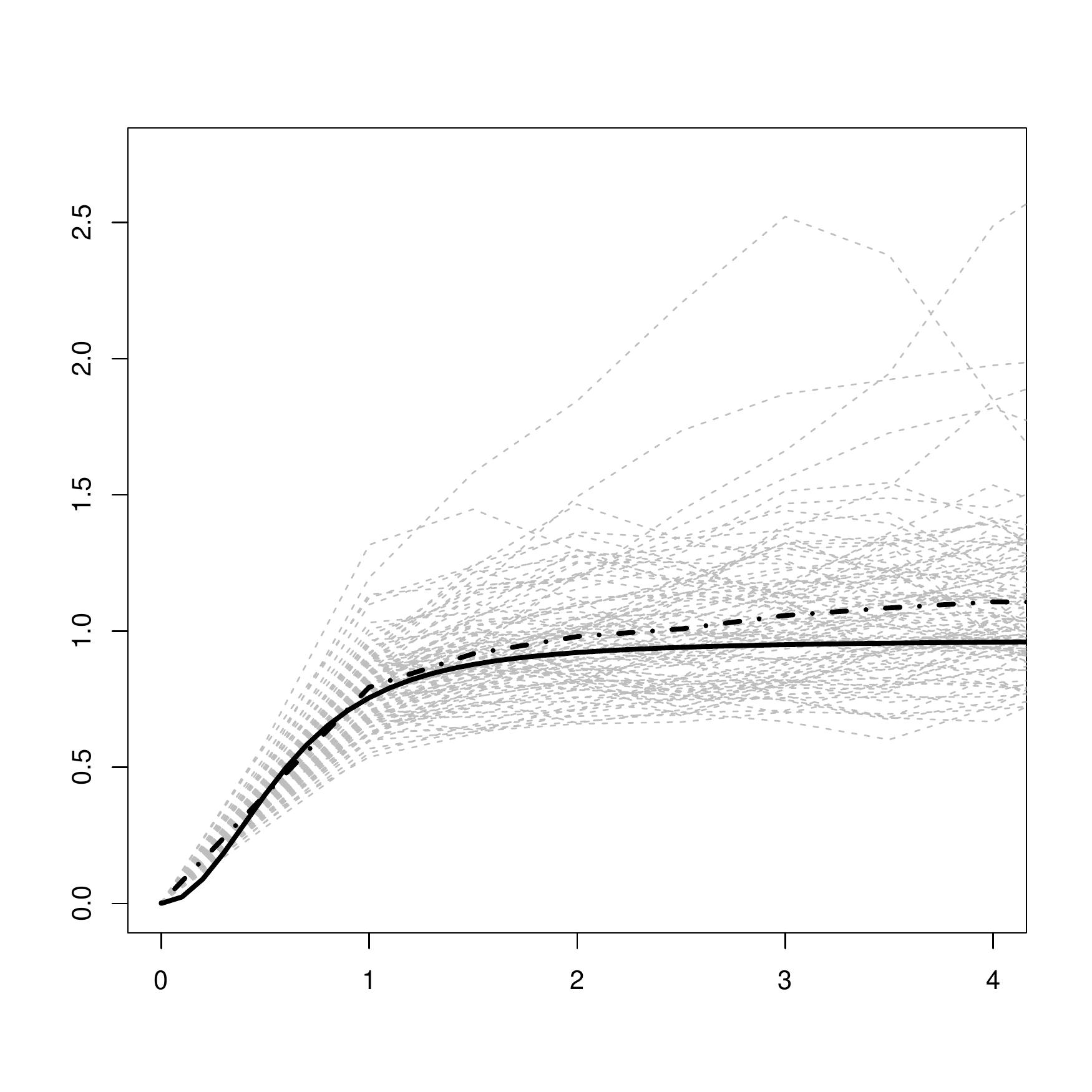} 
\caption{Temporal margin}
\label{gen_tm}
\end{subfigure}
\hfill
\begin{subfigure}{.32\linewidth}
\centering
\includegraphics[scale=0.3]{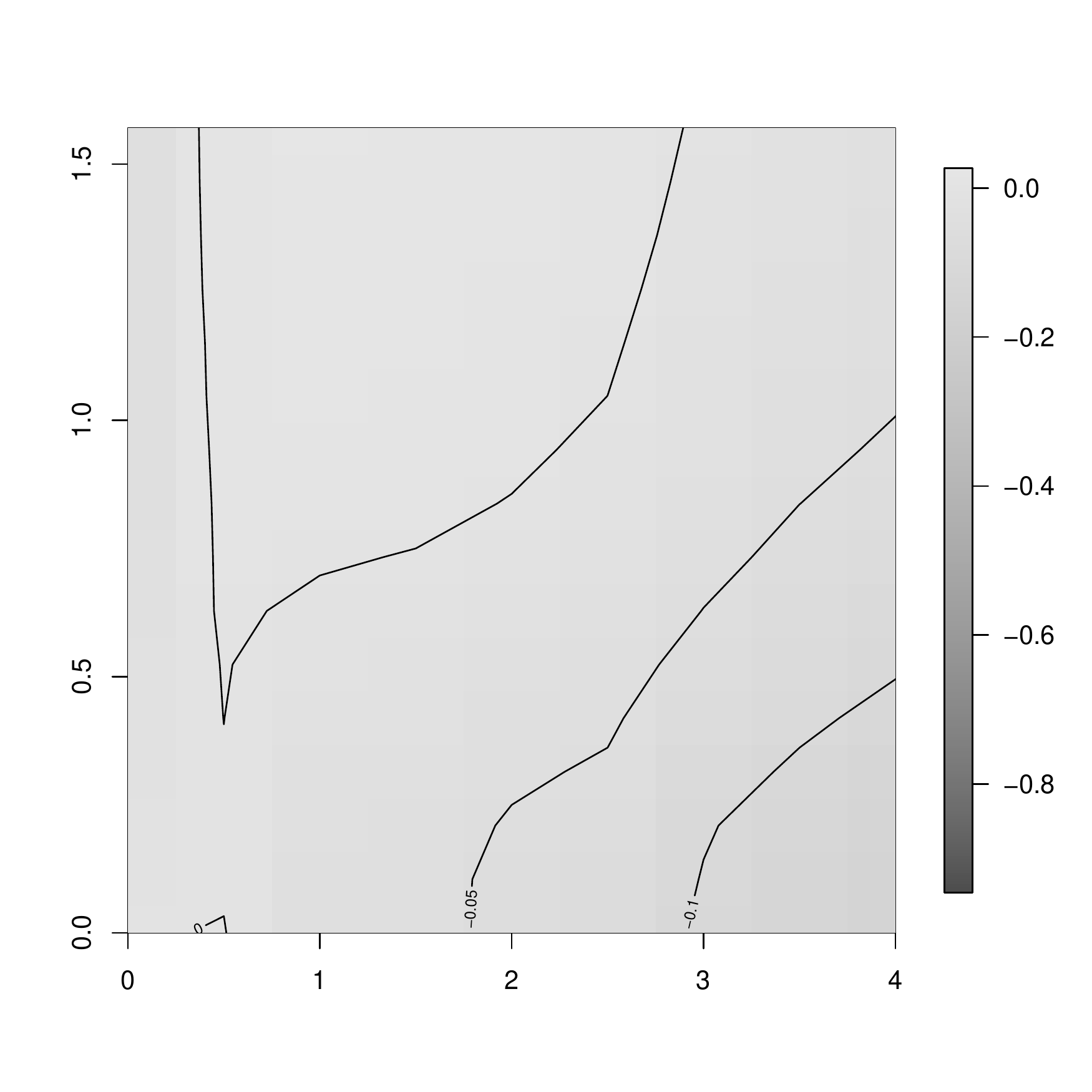} 
\caption{Variogram residuals}
\label{gen_vr}
\end{subfigure}
\caption{Plots of the spatial margin, temporal margin and residuals with respect to the mean variogram for $100$ simulated random fields. Panels (a) -- (c) and (d) -- (f) shows the behavior of the variogram when the true covariance function is given by \eqref{stcova1} using $g_{0} = \exp(-u/c_{0})$ and $g_{1} = (1+(\theta/c_{1})^{2})^{-1}$ respectively. The black line is the true variogram function, gray lines are the estimated variogram function for each simulation and black dotdashed points are the mean of the simulations.}\label{ST_variog_2}
\end{figure}

\noindent We simulate using the spherical grid $M = 30, N = 60$ and the temporal grid $\tau = \{1, 3/2, \ldots, T-1/2, T \}$ where $T = 8$, that is, $n = 28800$. We use the following family of covariance functions \citep{Porcu2016}:

\begin{equation}\label{stcova1}
C_{i}(\theta,u) = \left(\frac{1 - \delta}{1 - \delta g_{i}(u)\cos(\theta) }\right)^{\tau}, \quad \theta \in [0,\pi], u \in \R, i = 0,1
\end{equation} 

\noindent where $\delta \in (0,1)$, $\tau > 0$ and $g_{i}(u)$ is any temporal covariance function. In this case we consider $g_{0}(u) = \exp(-u/c_{0})$ and $g_{1}(u) = (1+(u/c_{1})^{2})^{-1}$. We set $\delta = 0.95, \tau = 1/4$, $c_{0} = 1.8951$ and $c_{1} = 1.5250$. Such setting ensures that $0.0470 < C_{0}(\theta,3) = C_{1}(\theta,3) < 0.0520$ for $\theta \in [0,\pi]$. Following \citet{Porcu2016}, Equation \eqref{stcova1} is a positive semi-definite covariance function, and so we use SVD decomposition to compute $\Lambda^{1/2}$. In addition to Table \ref{TableTime}, the number of points used in this experiment does not allow to use Cholesky decomposition, and so envelopes were not computed this time. Figure \ref{ST_variog_2} shows the estimated variogram for 100 simulations, concluding that we are simulating from the wanted distribution.

Finally, we use the method to simulate a spatio-temporal process with $N = 180$, $M = 360$, $\tau = \{0, 0.1, \ldots, 7.9, 8.0\}$ where $T = 8$, that is, $n = 5,248,800$, for $C_{1}(\theta,u)$ and $C_{2}(\theta,u)$ respectively. Such realizations are shown in Movie~1 and Movie~2 respectively.

\section{Discussion}

Circulant embedding technique was developed on $\S^{2}$ and $\S^{2} \times \R$ for an isotropic covariance function and for a spatially isotropic and temporally stationary covariance function respectively. All the calculations were done using the geodesic distance on the sphere. However, this method can be used with the chordal distance and axially symmetric covariance functions \citep{huang2012simplified}.
As shown in our simulation study, this method allows to simulate seamlessly up to $5.10^6$ points in a spatio-temporal context. Traditional functional summary statistics, like the variogram, require a high computational cost which motivates the development of different functional summary statistics or algorithms than can deal with a huge number of points.


Extensions to non-regular grids could be done using the technique detailed in \citet{dietrich1996fast}, which could be easily modified to $\S^{2}$ and $\S^{2}\times\R$. In addition, for an integer $k$, $\Sigma^{k}$ can be computed using circulant embedding by computing $\bLambda^{k}$ or $\bUpsilon^{k}$. Such result is also useful to compute $\Sigma^{k}$ with $k = -1.$ Such case corresponds to the inverse of a matrix \citep{jun2008nonstationary} which is important for the computation of maximum likelihood estimators and Kriging predictors \citep{stein2012interpolation}.

\section{Acknowledgments}
First author was supported by The Danish Council for Independent Research | Natural Sciences, grant DFF – 7014-00074 "Statistics for point processes in space and beyond", and by the "Centre for Stochastic Geometry and
Advanced Bioimaging", funded by grant 8721 from the Villum Foundation. Third author was supported by FONDECYT number 1170290.










\label{fig:sim_sphere}

\begin{figure}
\centering
\begin{subfigure}{0.5\linewidth}
\centering
\includegraphics[scale=0.4]{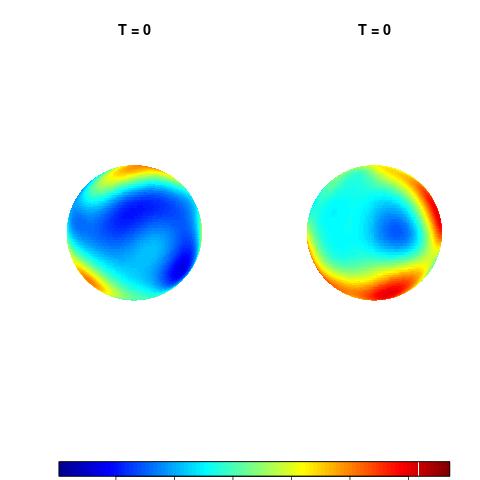}
\caption{Movie 1}
\end{subfigure}\\
\begin{subfigure}{0.5\linewidth}
\centering
\includegraphics[scale=0.4]{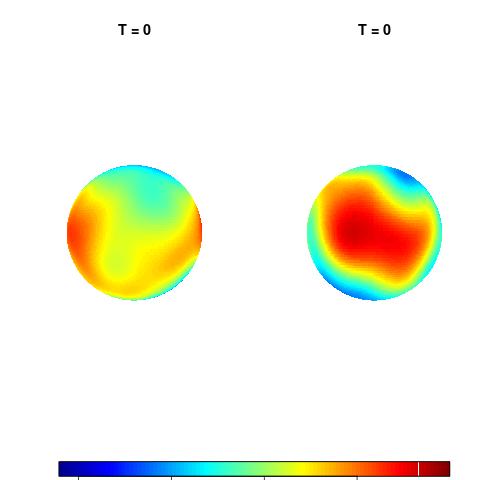}
\caption{Movie 2}
\end{subfigure}
\caption{Movie 1 and 2 show a realization of a spatio-temporal GRF with covariance function given by \eqref{stcova1} using $g_{0} = \exp(-u/c_{0})$ and $g_{1} = (1+(\theta/c_{1})^{2})^{-1}$ respectively.}
\end{figure}

\clearpage

\newpage

\bibliographystyle{apalike}
\bibliography{bibliografia.bib}

\end{document}